\pdfoutput=1
\documentclass[final,3p,times,twocolumn]{elsarticle}
\usepackage{amssymb}
\usepackage{lineno,hyperref}
\usepackage{subfigure}
\usepackage{siunitx}
\usepackage{booktabs}
\usepackage{relsize}

\def\textsubscript#1{\ensuremath{_{\mbox{\textscale{.6}{#1}}}}}
\def\textsuperscript#1{\ensuremath{^{\mbox{\textscale{.6}{#1}}}}}
 
\journal{Materials Science and Engineering: A}

\begin{document}

\begin{frontmatter}

\title{Modified \emph{Z}-Phase Formation in a 12\% Cr Tempered Martensite Ferritic Steel during Long-Term Creep}


\author{Johan Ewald Westraadt\corref{cor1}}
\ead{johan.westraadt@mandela.ac.za}
\cortext[cor1]{Corresponding author}
\author{William Edward Goosen}
\address{Centre for HRTEM, Nelson Mandela University, Box 77000, Gqeberha, 6031, South Africa}

\author{Aleksander Kostka}
\address{Centre for Interface-Dominated High Performance Materials (ZGH), Ruhr-University, Bochum, D-44801, Germany}

\author{Hongcai Wang}
\address{Institute for Materials, Ruhr-University, Bochum, D-44801, Germany}

\author{Gunther Eggeler}
\address{Centre for Interface-Dominated High Performance Materials (ZGH), Ruhr-University, Bochum, D-44801, Germany}

\begin{abstract}
The formation of modified \emph{Z}-phase in a 12Cr1MoV (German grade: X20) tempered martensite ferritic (TMF) steel subjected to interrupted long-term creep-testing at \SI{550}{\celsius} and 120 MPa was investigated. Quantitative volumetric measurements collected from thin-foil and extraction replica samples showed that modified \emph{Z}-phase precipitated in both the uniformly-elongated gauge ($f_v$: 0.23 $\pm$ 0.02 \%) and thread regions ($f_v$: 0.06 $\pm$ 0.01 \%) of the sample that ruptured after 139 kh. The formation of modified \emph{Z}-phase was accompanied by a progressive dissolution of MX precipitates, which decreased from ($f_v$: 0.16 $\pm$ 0.02\%) for the initial state to ($f_v$: 0.03 $\pm$\ 0.01\%) in the uniformly-elongated gauge section of the sample tested to failure. The interparticle spacing of the creep-strengthening MX particles increased from ($\lambda_{3D}$: 0.55 $\pm$\ 0.05 $\mu m$) in the initial state to ($\lambda_{3D}$: 1.01 $\pm$\ 0.10 $\mu m$) for the uniformly-elongated gauge section of the ruptured sample, while the thread region had an interparticle spacing of ($\lambda_{3D}$: 0.60 $\pm$\ 0.05 $\mu m$). The locally deformed fracture region had an increased phase fraction of modified \emph{Z}-phase ($f_v$: 0.40 $\pm$ 0.20\%), which implies that localised creep-strain strongly promotes the formation of modified \emph{Z}-phase. The modified \emph{Z}-phase precipitates did not form only on prior-austenite grain boundaries and formed throughout the tempered martensite ferritic grain structure.

\end{abstract}

\begin{graphicalabstract}
\includegraphics[width=\textwidth]{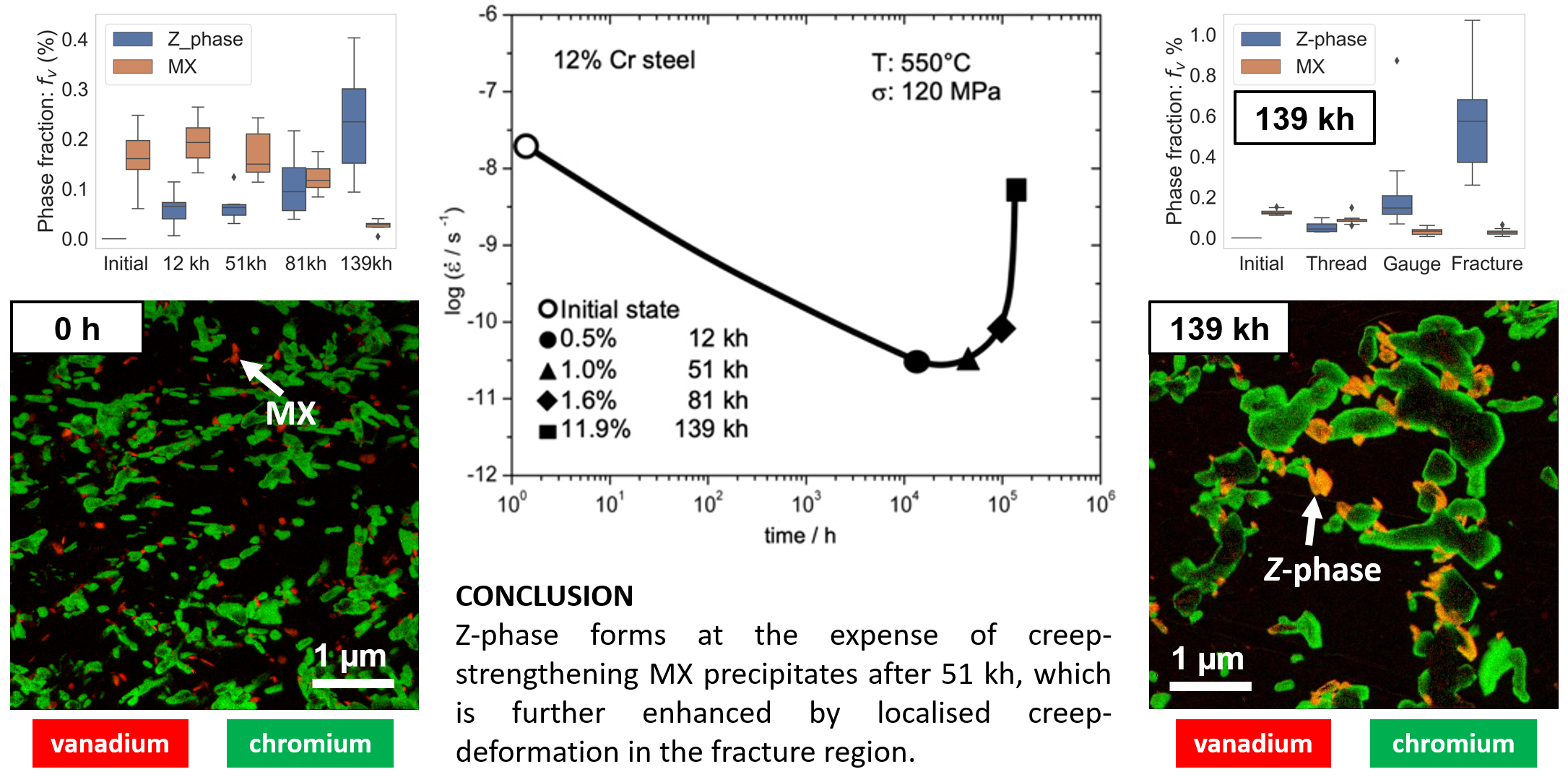}
\end{graphicalabstract}

\begin{highlights}
\item Formation of modified \emph{Z}-phase in a 12Cr1MoV steel during long-term  interrupted creep-testing up to 139 kh was quantitatively evaluated.     
\item Improved methods to estimate the  quantitative volumetric precipitate measurements from extraction replicas were applied. 
\item Modified \emph{Z}-phase was first identified in gauge section of the sample tested to 51 kh using electron diffraction. 
\item Applied stress and localised creep-strain (deformation) promoted the formation of modified \emph{Z}-phase.   
\end{highlights}

\begin{keyword}
  Tempered martensite ferritic steels \sep modified \emph{Z}-phase \sep long term creep \sep energy filtered TEM 
\end{keyword}

\end{frontmatter}


\section{Introduction}
Tempered martensite ferritic (TMF) steels with 9-12\%Cr are used extensively for steam pipes, turbines and boilers in fossil fired steam power plants. These steels have high creep strength, good oxidation resistance, good thermal fatigue properties and a moderate cost. The most common commercial material grades are 9Cr1Mo grade 91 (T/P91) or 9Cr2W grade 92 (T/P92), but older 12Cr1MoV (German grade: X20) is currently still in operation at several power utilities \citep{vanrooyen2019}.

The TMF steel structure consists of a hierarchical arrangement of prior-austenite grains (PAG), packets, blocks and micro-grains with a high density of free dislocations \citep{aghajani2009b, kostka2007, sawada2011, pesicka2003, pesicka2010, payton2012}. TMF steels are stable during creep-exposure due to the pinning forces exerted by particles of different origin. M\textsubscript{23}C\textsubscript{6} carbides, located at grain boundaries, and MX carbonitrides, distributed homogeneously throughout the ferrite matrix, acts as pinning agents, by impeding free-dislocation movement and suppresses the movement of grain boundaries during creep \citep{kostka2007, abe2009}. During creep conditions, coarsening of M\textsubscript{23}C\textsubscript{6} and Laves phase and dissolution of MX carbonitrides occurs, which decreases the pinning forces exerted by the particles. This promotes the evolution of low angle grain boundaries, which is considered to be the most important strengthening mechanisms of 9-12\%Cr TMF steels, leading to a decrease in the long-range internal stress field \citep{abe2009, meier1993, blum2009}. 

New generation TMF steels are typically designed with an 11-12 wt.\% chrome content to have sufficient oxidation resistance at the higher operation temperatures (\SI{650}{\celsius}) needed for improved power plant efficiency. Although, short-term testing on these new generation steels (up to 10 kh) has shown an enhancement in creep-strength and oxidation resistance, long-term testing (longer than 50 kh) revealed microstructural instabilities resulting in a creep-strength break-down \citep{danielsen2016}. Extensive microstructural investigations performed by Strang and Vodarek \citep{vodarek2003}, Danielsen \citep{danielsen2006a, danielsen2007} and Sawada \citep{sawada2006, sawada2014} on service exposed and lab-tested TMF steels have shown that this creep-strength breakdown, in 11-12\%Cr TMF steels, is associated with the transformation of fine MX carbonitrides into larger modified \emph{Z}-phase particles, which results in an increase in the inter-particle spacing. In addition, the interface between coarse modified \emph{Z}-phase and the iron matrix could serve as a nucleation site for creep voids. In contrast, the creep-strength breakdown in 9\% Cr TMF steels is not considered to be related to the MX to Z-phase transformation \citep{fedoseeva2016, fedoseeva2016a, yoshizawa2009, sawada2008}.  

\subsection{Modified \emph{Z}-Phase formation in 9-12\%Cr TMF steels}
The formation of Cr(Nb,V)N modified \emph{Z}-phase in 9-12\% Cr TMF steels has been the subject of numerous studies \citep{strang1996, danielsen2006a, sawada2006, zhou2015, fedoseeva2016}. Danielsen \citep{danielsen2006a} gives an excellent historical review of modified \emph{Z}-phase precipitation in steels and investigated ten 9-12\%Cr service exposed TMF steels for the presence of modified \emph{Z}-phase. A more recent review of the modified \emph{Z}-phase behaviour in 9-12\%Cr steels by Danielsen \citep{danielsen2016} was made in reference to the design of next-generation steels to overcome this microstructural instability. 

The modified \emph{Z}-phase has a chemical composition close to 1/3Cr, 1/3(V+Nb) and 1/3N, where the V/Nb ratio can vary significantly for precipitates in a particular steel. Two types of modified \emph{Z}-phase are found in 9-12\% Cr TMF steels: (1) a metastable fcc phase with a lattice parameter of 0.405 nm, and (2) a stable tetragonal phase which is a distortion of the NaCl-type lattice. Electron diffraction measurements of Cr(V,Nb)N and CrVN particles showed that both cubic and tetragonal diffraction patterns could be obtained from the same particles \citep{danielsen2006, danielsen2009}. This dual cubic/tetragonal crystal structure is believed to be connected to the transformation process, with the cubic crystal structure being a metastable step towards the more stable tetragonal crystal structure. Two types of nucleation mechanisms are reported for modified \emph{Z}-phase particles in 9-12\% Cr TMF steels. The most commonly observed mechanism, involves direct transformation of MX carbonitrides into modified \emph{Z}-phase by diffusion of chromium, which leads to the formation of coarse particles. Nucleation of modified \emph{Z}-phase on the surface of MX carbonitrides is rarely observed and leads to the formation of nanoscale modified \emph{Z}-phase particles.  

Thermodynamical calculations have shown that matrix concentration  of chromium is the most influential element driving the formation of modified \emph{Z}-phase \citep{danielsen2007a}. The nitrogen content also increases the formation of modified \emph{Z}-phase since it is a relatively pure nitride compared to the stable MX carbo-nitrides that can form in relatively low nitrogen environments. MX particles consisting of Nb-rich (Nb,V)N were found to preferentially transform into modified \emph{Z}-phase during high-temperature exposure \citep{cipolla2010b}. This result is consistent with the analysis performed on an 12\%Cr Nb-free X20 steel grade, which show relatively slow precipitation of CrVN modified \emph{Z}-phase compared to the Nb-containing steel grades \citep{danielsen2006a}. It was then suggested that new generation steels with increased chromium content (11\% Cr) for oxidation resistance should be developed with a lowered niobium content to slow down modified \emph{Z}-phase formation. A new generation steel (THOR115) was developed based on this principle and has shown good microstructural stability for creep-testing up to 60 kh \citep{ortolani2017}, but longer-term testing is still in progress. 

Previous experimental observations based on extraction replicas found that the modified \emph{Z}-phase precipitated preferentially along prior-austenite and packet grain boundaries \citep{sawada2006, cipolla2010a, sawada2014}. Investigations performed on ruptured creep-tested samples have shown that the gauge portions of the ruptured samples contained 2-4 times the number density of modified \emph{Z}-phase particles compared to the thread portions of the samples after creep-rupture \citep{sawada2006, sawada2014}. These investigations were performed on the locally deformed (necking) sections of the gauge area, which were subjected to long-term stress/temperature as well as creep-strain deformation. Recent thermodynamical kinetic precipitate modelling that includes the effects of the volumetric misfit across the precipitate/matrix interface, has shown that this misfit may influence the transformation and growth kinetics of modified \emph{Z}-phase significantly \citep{svoboda2019}. 

\subsection{Microstructural evolution of 12Cr1MoV (German grade: X20) during long-term creep}
X20 is an older German steel grade based on 12\%Cr with a relatively high carbon content of 0.2 wt\%. The chemical composition of the steel used in this study can be seen in Table \ref{table:chemistry}, which includes measurements for the N and Nb content, which are not typically controlled during manufacturing, but are considered to be important in the formation of modified \emph{Z}-phase.  

The presence of modified \emph{Z}-phase in X20 grade steels was first observed by Danielsen \citep{danielsen2006} in the analysis of a service exposed sample (150 kh at \SI{600}{\celsius}). A few particles of modified \emph{Z}-phase (CrVN), exhibiting a cubic crystal structure (a\textsubscript{0}=\SI{0.405}{\nm}) were identified with STEM-EDS and SAED analysis, while several MX particles were still observed in the sample. In comparison, they found that the Nb-containing steel grades consistently contained more Cr(Nb,V)N modified \emph{Z}-phases.

The microstructural evolution of the 12Cr1MoV TMF steel samples exposed to long-term interrupted creep-testing at near operating temperatures (\SI{550}{\celsius}) and 120 MPa stress has been the subject of several extensive investigations. The gauge and thread portions of the interrupted creep-tested samples have been characterised to quantify precipitates \citep{eggeler1989a, aghajani2009b, aghajani2009}, dislocations \citep{pesicka2003, pesicka2010}, and sub-grains \citep{payton2012} previously. The previous precipitate analyses were performed on thin-foils using HAADF-STEM combined with EDS and SAED analysis to determine the compositional and crystallographic information. Modified \emph{Z}-phase precipitation was not observed in any of the samples and the MX precipitate population was found to be stable up to 139 kh \citep{aghajani2009b}. However, recent experimental work \citep{wang2021} based on EFTEM analysis performed on extraction replicas prepared from the samples of the previous investigation showed clear evidence of modified \emph{Z}-phase formation in the uniformly-elongated gauge section after 51 kh of creep-testing. Recent studies \citep{marx2019} conducted on ex-service 12Cr1MoV material grades exposed to long-term operation, show clear evidence of modified \emph{Z}-phase formation, but this microstructural instability has not yet been the subject of a systematic investigation for this steel grade.

\subsection{Quantitative characterisation of modified \emph{Z}-phase}
Investigations of modified \emph{Z}-phase precipitation in 9-12\%Cr TMF steels have been mainly performed on \emph{extraction replicas} using TEM-based techniques (SAED, EFTEM and STEM-EDS/EELS). Extraction replicas are relatively easy to prepare and provide large electron transparent areas available for analysis of the precipitate phases, without the influence of the magnetic matrix. However, the sampled volume is unknown and quantitative results are often reported as the projected area measurements to quantify the phase fraction ($f_A$) and number density ($N_A$) of precipitates \citep{sawada2010, sawada2011}. \emph{Bulk} SEM based methods can also be used to quantify the precipitate species in 9-12\%Cr TMF steels \citep{byrne2020}, but the quantification of the fine MX carbonitrides are not possible due to insufficient spatial resolution. \emph {Thin-foils} have been used extensively to study precipitates in 9-12\%Cr CSEF steels \citep{sawada2014, eggeler1989a, aghajani2009}.  The main advantage of thin-foils is that the location specific information of the precipitates are preserved, the analysed volume can be determined by measuring the thickness of the foil, and the spatial resolutions in the TEM is sufficient to resolve the very fine MX precipitates. However, the available sample volumes are limited to the electron transparent areas of the foil.

The present experimental study applies volumetric quantification methods to systematically study the formation of CrVN modified \emph{Z}-phase and the associated dissolution of the MX precipitates during long-term creep in a 12Cr1MoV steel. In addition, results on the reaction kinetics (time), effects of stress, localised creep-deformation, and the preferential nucleation sites of the modified \emph{Z}-phase are presented.

\section{Materials and Methods}
The TMF steel investigated in this study was supplied by Salzgitter Mannesmann Research Centre, Duisburg. The chemical composition given in Table \ref{table:chemistry} was measured using optical emission spectroscopy from the thread section of the sample tested to rupture. The steel was austenitized at \SI{1050}{\celsius} for 30 minutes followed by air-cooling to room temperature. The steel was then tempered at \SI{770}{\celsius} for 2 hours and allowed to air-cool to room temperature. 

\begin{table*}[!ht]
  \caption{Chemical composition (in mass\%) of the steel grade.}
  \small
  \centering
  \label{table:chemistry}
  \begin{tabular}{l c c c c c c c c c c c c}
    \toprule
    Grade & C & Si & Mn & S & Ni & Cr & Mo & V & Nb & N & Al & Fe \\ 
    \midrule
    X20 & 0.21 & 0.15 & 0.63 & 0.010 & 0.76 & 12.1 & 1.00 & 0.24 & $\leq$0.004 & 0.044 & 0.006 & bal. \\
    \bottomrule
  \end{tabular}
\end{table*}
 
\begin{figure}
  \begin{center}
    \includegraphics[width=0.45\textwidth]{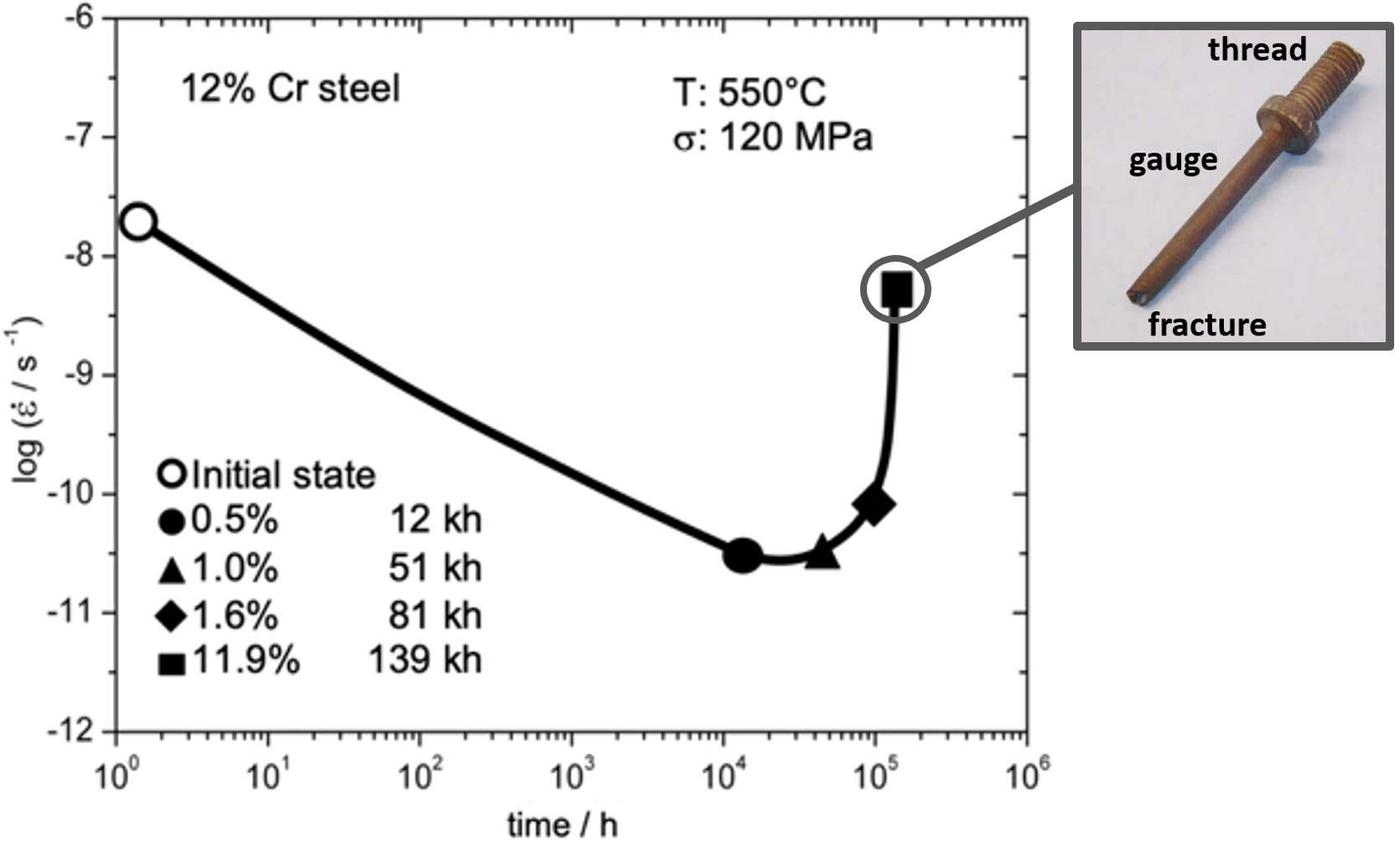}
  \end{center}
  \caption{Material states investigated in this study. The insert shows the sample tested to failure, which fractured after 139 kh, indicating the different areas that were investigated. This includes the \emph{thread} region, uniformly-elongated \emph{gauge} region, and \emph{fracture} region which experienced localised creep-deformation.}
  \label{fig:creep_curve}
  \end{figure}

Four identical samples were tested simultaneously at \SI{550}{\celsius} with an applied stress of 120 MPa. Three interrupted experiments were tested to creep-strain values of 0.5\% (12 kh), 1.0\% (51 kh), and 1.6\% (81 kh) respectively, and one sample was taken to rupture at a creep-strain value of 11.5\% after 139 kh. Figure \ref{fig:creep_curve} shows the interrupted material states that were investigated. For each creep-tested sample the gauge and thread regions were investigated to study the effects of stress (120 MPa) on the microstructure. The fractured region, which experienced localised creep-deformation, was also investigated to study the effects of the tertiary creep-deformation on the microstructure.  

  Thin-foils were prepared from the uniformly-elongated gauge and thread sections of the creep-tested samples for each of the material states. In addition, thin-foils were prepared from a reference X20 steel in the as-tempered state. Thin slices were cut from the sample, mechanically thinned down to \SI{300}{\mu m}, punched into 3 mm disks, and mechanically thinned down to \SI{80}{\mu m} in thickness. The disks were then electropolished to electron transparency using a 5\% perchloric acid/methanol solution at \SI{-20}{\celsius} and 21 V with a Struers Tenupol 5 twin-jet electro-polisher. Extraction replicas were prepared from the same set of material states, by first etching the surface for 30 seconds with Viella's solution, depositing a 30 nm thick layer of carbon, scoring the surface into 1 mm blocks and then placing it back in the solution. The liberated carbon films contain the precipitates in the steel, which are then collected, washed with a methanol/ethanol solution, then placed onto a copper TEM grid, and allowed to air dry. FIB lamellae were prepared from ten sites in the fracture region in the ruptured sample (139 kh) using a Ga-ion FIB-SEM. Extraction replicas were also prepared from this fracture region.   

  The chemical composition of the precipitates was investigated using scanning transmission electron microscopy (STEM) combined with energy dispersive X-ray spectroscopy (EDX) and electron energy loss spectroscopy (EELS) on a JEOL2100 (LaB\textsubscript{6}) TEM fitted with an Oxford X-Max (80 mm\textsuperscript{2}) X-ray spectrometer and a Gatan Quantum GIF electron energy loss spectrometer. The crystallographic information of the precipitates was determined using selected area electron diffraction (SAED) patterns taken from several different zone-axes and matching it to theoretically simulated patterns. The precipitates (M\textsubscript{23}C\textsubscript{6}, MX and modified \emph{Z}-phase) were mapped on ten sites of interest using energy filtered TEM (EFTEM) (512x512 pixels; 5x5 $\mu m^2$) for each material state and sample preparation method (thin-foil and extraction replica). The signals for Cr (green) and V (red) were overlayed to produce a RGB composite image. M\textsubscript{23}C\textsubscript{6} precipitates rich in chromium appeared as green particles, while the vanadium rich MX precipitates appeared as red particles. Modified \emph{Z}-phase contains both chromium and vanadium and appeared as orange particles. A thickness map using the log-ratio method was collected on the thin-foil samples sites of interest, which was used to determine the sampled volume.  

  The composite RGB images were segmented into M\textsubscript{23}C\textsubscript{6}, MX and modified \emph{Z}-phase precipitates with the MIPAR \citep{sosa2014} image analysis software based on colour. The projected areas, centroid position and equivalent circle diameters for each precipitate were measured and exported to a text file. These centroid positions were then used to read the thickness of the foil from the thickness map. Stereological corrections described in \citep{sonderegger2006} were applied to the exported precipitate measurements to calculate the corrected values for the volumetric phase fraction ($f_v$),  mean precipitate size ($d_m$), and mean precipitate spacing ($\lambda_{3D}$) of a particular precipitate species.

  The corrections and calculations were performed separately for each site of interest, which resulted in ten measured values for each material state. This distribution of measurements was then plotted as a box plot to show the distribution of values obtained from each site of interest, which was then used to calculate the mean value and the standard error in the mean for a particular material state. Any data points falling outside 1.5 x interquartile range (IQR) were considered outliers for the box plots, but these outliers were still used to determine the quantitative measurements.  A total area of 10x25$\mu m^2$ was investigated for each material state. The total material volume investigated depends on the sample preparation method (thin-foil or extraction replica). The sampled volume can be calculated by multiplying with the foil thickness (80 nm) or the extraction depth (1 $\mu m$) in the case of the extraction replica with the projected area in the field of view. The extraction depth is unknown for the extraction replica preparation method, however, it can be estimated with the assumption that the M\textsubscript{23}C\textsubscript{6} precipitate phase fraction remains constant during long-term creep \citep{aghajani2009b}. The M\textsubscript{23}C\textsubscript{6} projected area phase fraction measurements from the extraction replicas were combined for the ten sites of interest for a particular material state and normalised to have a volumetric phase fraction of 4.5 \%, which was based on the chemical composition and equilibrium thermodynamic modelling. This was then used to calculate the total precipitate extraction volume which was then divided by the total area (250 $\mu m^2$) to estimate the average extraction depth for a particular material state. 

  SAED experiments are very time-consuming and difficult to perform in the case of magnetic samples. Transmission Kikuchi diffraction \citep{sneddon2016} provides an easy and accessible alternative method that can be routinely performed on most modern SEMs fitted with an EBSD detector. This method allows for crystallographic phase and orientation information to be collected with step sizes as small as 10 nm for electron transparent samples, which is a significant improvement over bulk EBSD methods. In addition, the diffraction data is automatically processed during the data acquisition and does not require additional indexing, as is the case with SAED. The samples were investigated using TKD combined with EDS in a JEOL7001F FEG-SEM fitted with a Nordlys HKL EBSD detector. Step sizes of 10-50 nm were used depending on the objective of the experiment at an accelerating voltage of 30 kV and a sample tilt of \SI{-20}{\degree}.

\section{Results}

\begin{figure}
  \begin{center}
    \includegraphics[width=0.45\textwidth]{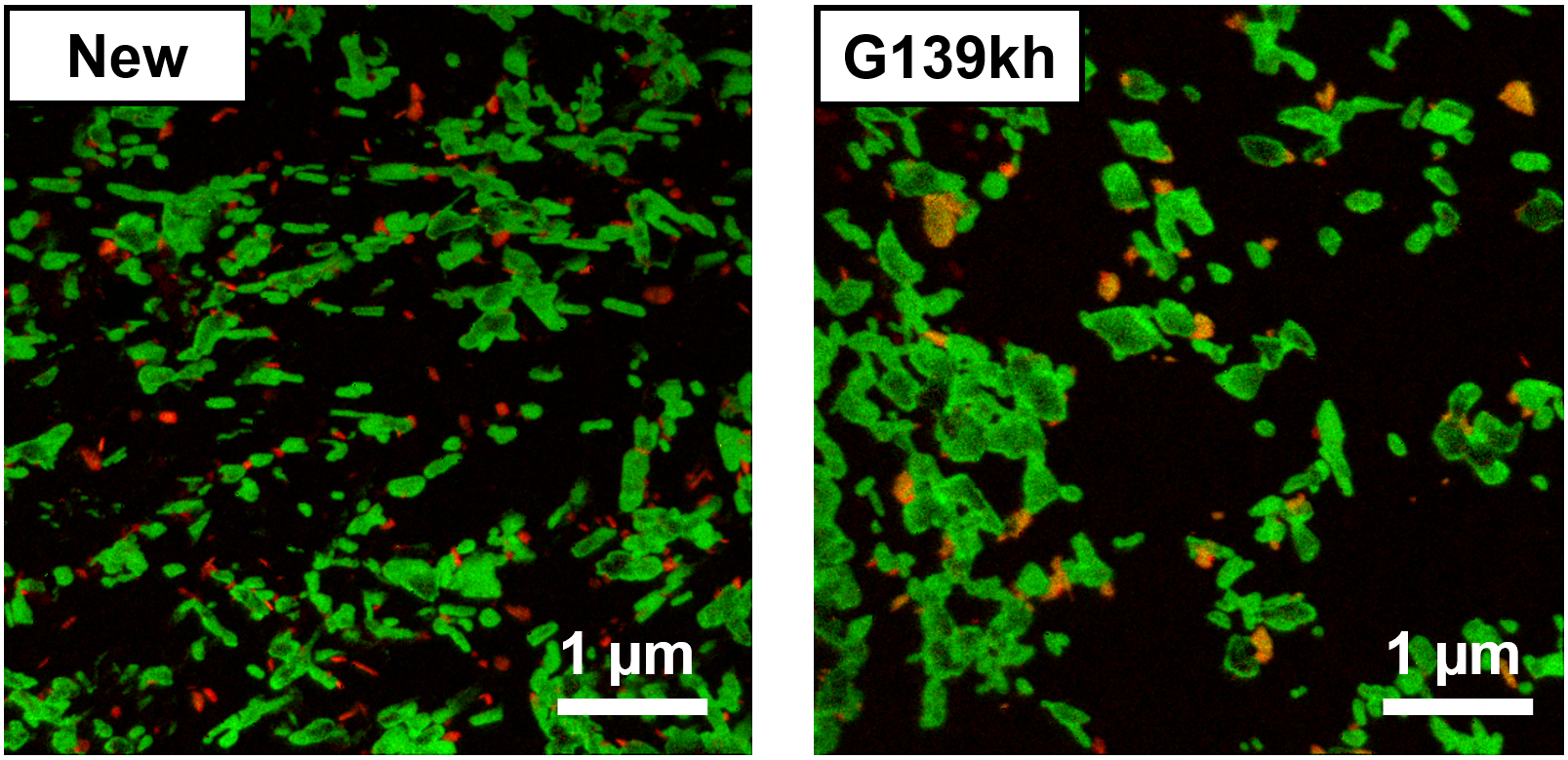}
  \end{center}
  \caption{EFTEM elemental maps shown as a colour overlay (Cr:green; V:red) for the starting material (New) and the  uniformly-elongated gauge section of the sample creep-tested to failure (G139kh) collected from extraction replicas.}
  \label{fig:EFTEM_RGB_Replica}
  \end{figure}

Figure \ref{fig:EFTEM_RGB_Replica} shows typical EFTEM RGB composite elemental maps taken from the replica samples prepared from the initial state and the uniformly-elongated gauge section of the ruptured sample. The initial state shows the Cr-rich M\textsubscript{23}C\textsubscript{6} particles in green and several smaller V-rich MX particles in red. The MX particles in the ruptured sample decreased in number density and several additional Cr-V orange particles, with a chemical composition consistent with the modified \emph{Z}-phase can be seen.

\subsection{Identification of the modified \emph{Z}-phase precipitates}
\begin{figure}
  \begin{center}
    \includegraphics[width=0.45\textwidth]{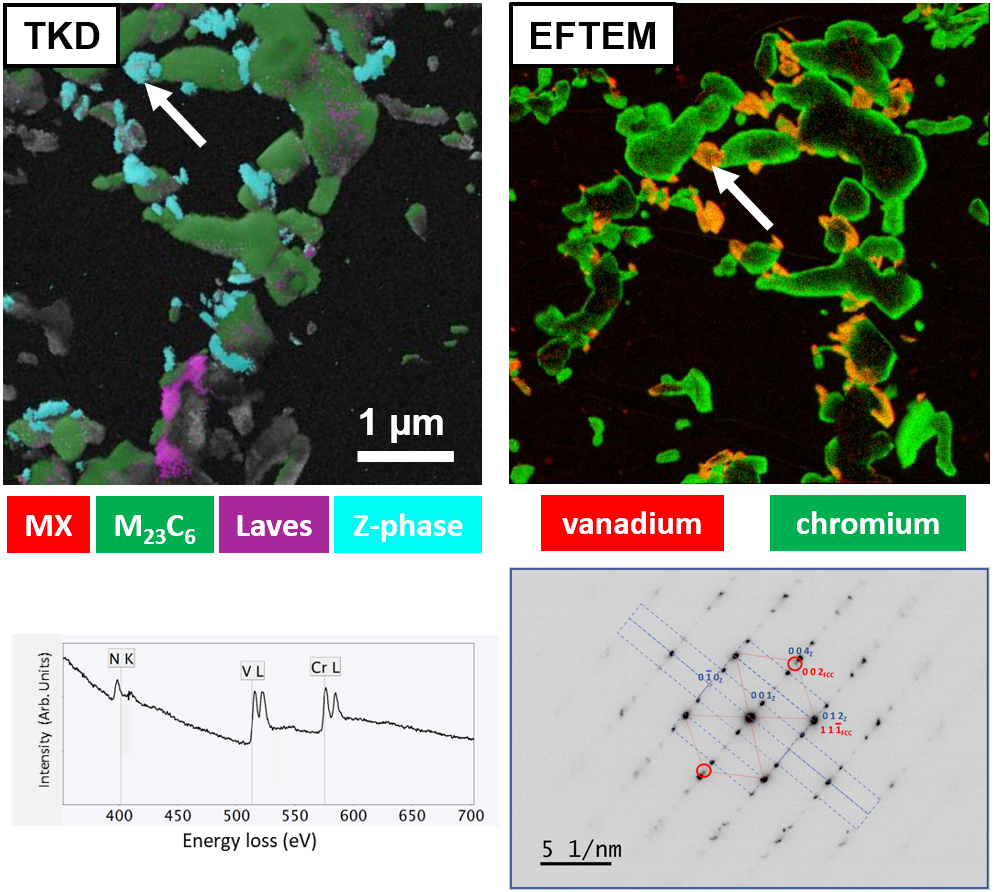}
  \end{center}
  \caption{(a) TKD crystallographic phase map showing the distribution of modified \emph{Z}-phase particles with a tetragonal crystal structure, (b) EFTEM elemental map overlay showing the modified \emph{Z}-phase particles in orange. A modified \emph{Z}-phase particle (white arrow) was analysed using (c) STEM-EELS indicating that the particle consists of Cr-V-N and (d) SAED on a \textbf{B} = [100] tetragonal zone axis, showing the reflections of the tetragonal phase (Z-blue) and reflections of the cubic variant \textbf{B} = [101] (FCC-red), the 200-type reflections are indicated with red circles.}
  \label{fig:ZPhaseID}
  \end{figure}

The identification of modified \emph{Z}-phase/MX/M\textsubscript{2}X precipitates based only on EFTEM maps must be done with care, since all of these particles contain chromium and nitrogen. The identification of the modified \emph{Z}-phase based on the EFTEM RGB composite elemental map overlay was confirmed using diffraction (SAED and TKD) and compositional (STEM-EDS/EELS) measurements. Figure \ref{fig:ZPhaseID} shows a TKD crystallographic phase map and the corresponding EFTEM RGB composite elemental map taken from an extraction replica prepared from the uniformly-elongated gauge section of the fractured sample. Both the cubic and tetragonal modified \emph{Z}-phase as well as M\textsubscript{23}C\textsubscript{6}, Laves phase, and MX candidate crystal structures were provided to the indexing software. The energy loss spectrum (bottom left insert) shows that the precipitate consisted of Cr-V-N, with EDS analysis (not shown) confirming that the Cr:V elemental ratio is close to unity. The SAED zone-axis pattern taken from this precipitate, exhibits both the cubic \textbf{B}=[101] reflections (red) and the additional tetragonal \textbf{B}=[100] (blue) reflections (insert bottom right), which was first observed by Danielsen \citep{danielsen2006}. Several additional zone-axis SAED patterns taken from the same precipitate (not shown), confirmed this observation.

\subsection{Quantification of the MX/modified \emph{Z}-phase precipitates}
\begin{figure}
  \begin{center}
    \includegraphics[width=0.45\textwidth]{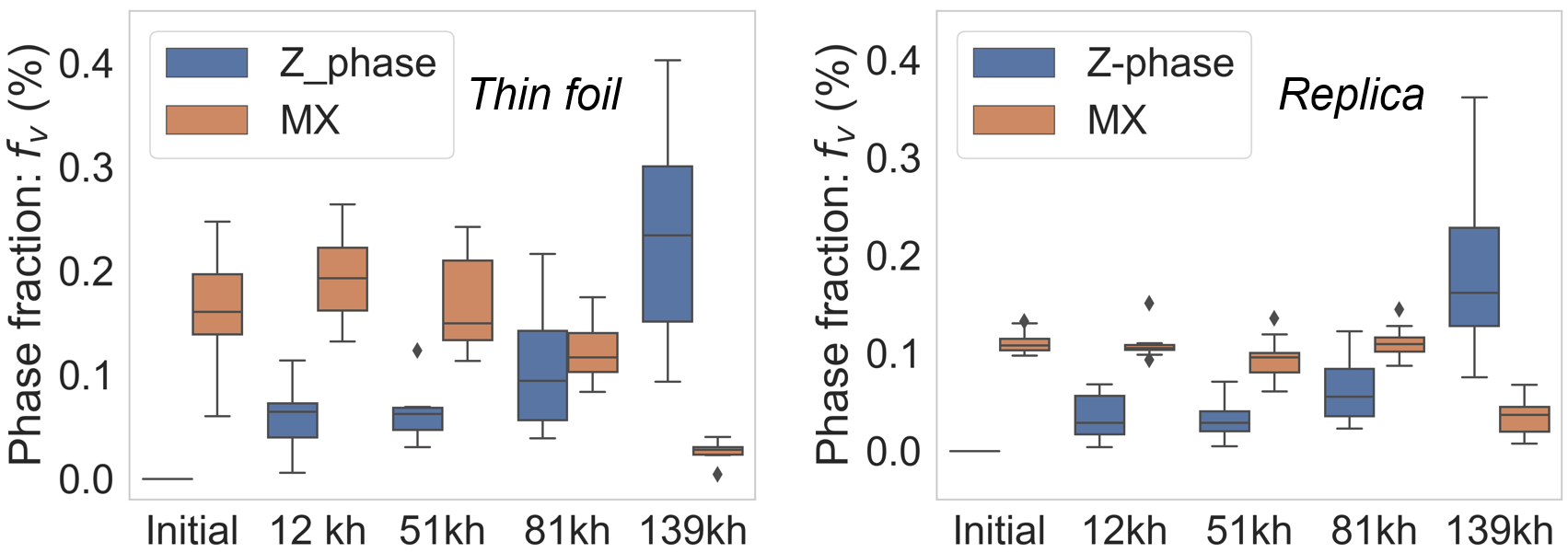}
  \end{center}
  \caption{Phase fraction (f\textsubscript{v}) of MX and modified \emph{Z}-phase precipitates in the gauge sections as a function of creep-testing time for data collected from the thin-foils and extraction replicas.}
    \label{fig:MX_Z_fv}
  \end{figure}

Figure \ref{fig:MX_Z_fv} shows the \emph{corrected} volumetric phase fraction ($f_v$) of the modified \emph{Z}-phase and MX precipitates for the uniformly-elongated gauge sections of the creep-tested samples and the initial state, as measured on the thin-foils (left) and extraction replicas (right). The box plots show the distribution of the mean values calculated for each of the ten sites of interest. Quantitative measurements based on TEM analyses should be done with care, due to the small volumes of material that are investigated. If the number density of features are high enough such that the field of view is representative of the bulk, then the standard deviation of the mean values of measurements taken from each site of interest should be small. The standard deviation for the measured phase fraction ($f_v$) for the MX precipitates are much less for the measurements taken from the extraction replicas, due to the larger sampling volume of this method. The phase fraction of MX precipitates is relatively stable ($f_v$ = 0.16 $\pm$ 0.02) to a test duration of 51 kh, but decreases sharply afterwards. This is accompanied by an increase in the phase fraction of modified \emph{Z}-phase. This result is consistent with the precipitation of modified \emph{Z}-phase and dissolution of the MX precipitates as observed by previous investigations by Sawada \citep{sawada2006} and Danielsen \citep{danielsen2006a}.

\begin{figure}
  \begin{center}
    \includegraphics[width=0.45\textwidth]{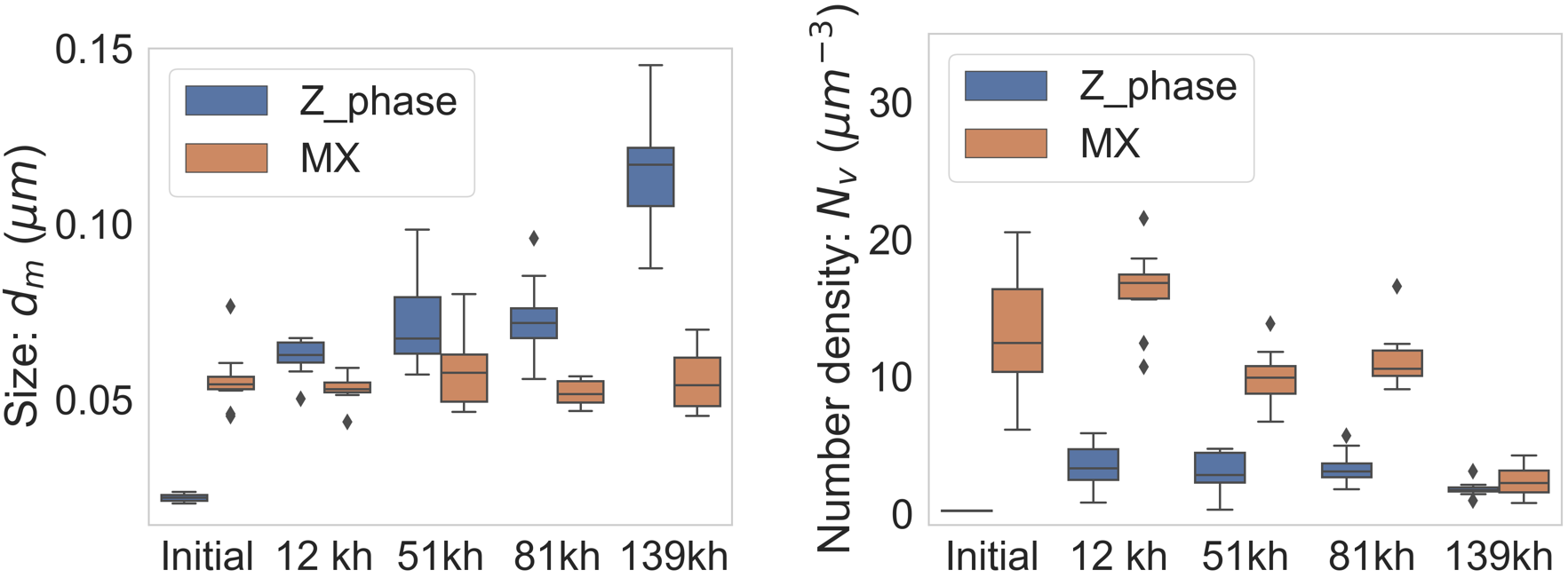}
  \end{center}
  \caption{Mean corrected size (d\textsubscript{m}: left) and corrected number density (N\textsubscript{V}: right) of MX and modified \emph{Z}-phase precipitates for different creep-testing times in the uniformly-elongated gauge regions for data collected from the thin-foil samples.}
  \label{fig:MX_Z_gauge_dm_Nv}
  \end{figure}

  Figure \ref{fig:MX_Z_gauge_dm_Nv} shows the distribution of the \emph{corrected} mean equivalent circle diameters (left) ($d_m$) and volumetric number density ($N_v$) (right) of the MX and modified \emph{Z}-phase precipitates based on measurements performed on the thin-foils prepared from the uniformly-elongated gauge sections and the initial state material. The size of the MX precipitates remained relatively constant during long-term creep-testing and with sizes ranging between $d_m$ = 50-60 nm. This result is in agreement with previous studies \citep{aghajani2009b}. The size of the modified \emph{Z}-phase precipitates increases from starting values similar to MX precipitates in the early stages (51kh) to $d_m$ = 0.113 $\pm$ 0.001 $\mu m$ for the modified \emph{Z}-phase precipitates observed in the  uniformly-elongated gauge section tested to rupture. The number density of the MX particles decreases sharply from $N_v$ = 16 $\pm$ 2 $\mu m^{-3}$ in the initial state to $N_v$ = 2 $\pm$ 1 $\mu m^{-3}$ for the ruptured sample. The number density of the modified \emph{Z}-phase remains relatively constant at $N_v$ = 3 $\pm$ 1 $\mu m^{-3}$.

\begin{figure}
  \begin{center}
    \includegraphics[width=0.45\textwidth]{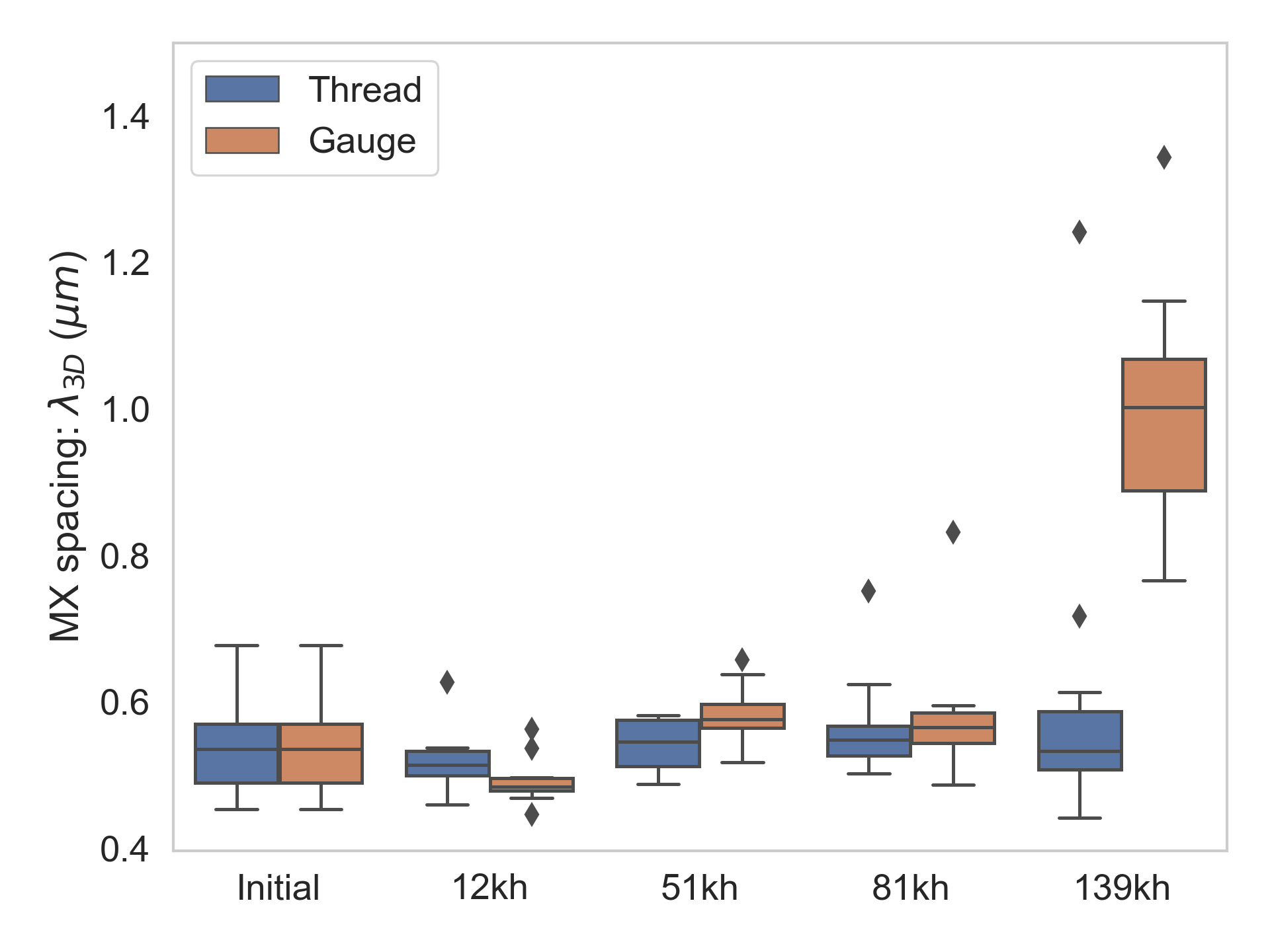}
  \end{center}
  \caption{Mean 3D centroid-centroid distance between the MX precipitates for the different material states for data collected from the thin-foil samples. Outliers are data points falling outside 1.5*IQR of the 25th and 75th percentiles.}
  \label{fig:MX_3D_lambda}
  \end{figure}

  Figure \ref{fig:MX_3D_lambda}  shows the calculated \citep{holzer2010a} average 3D centroid-centroid spacing ($\lambda_{3D}$) between the MX particles as a function of creep-testing time in the uniformly-elongated gauge and the thread portions of the creep-tested samples, based on measurements performed on the thin-foils. The results from the thread sections show the effect of thermal ageing of the material at a temperature of \SI{550}{\celsius}, while the results from the (uniformly-elongated) gauge sections include the effects of the 120 MPa stress that was applied. The average MX spacing for the initial material state is $\lambda_{3D}$ = 0.55 $\pm$ 0.05 $\mu m$ and only increases slightly for creep-testing to 81 kh for both the gauge and thread sections, which were very similar in value. There is, however, a large difference between the MX precipitate spacing in the  uniformly-elongated gauge section ($\lambda_{3D} = 1.01 \pm 0.10 \mu m$) as compared to the thread section ($\lambda_{3D} = 0.60 \pm 0.10 \mu m$) for the sample tested to failure, which implies that  creep-strain and/or stress promotes the formation of modified \emph{Z}-phase.

\subsection{Effect of creep-strain on the formation of modified \emph{Z}-phase}
\begin{figure}
  \begin{center}
    \includegraphics[width=0.45\textwidth]{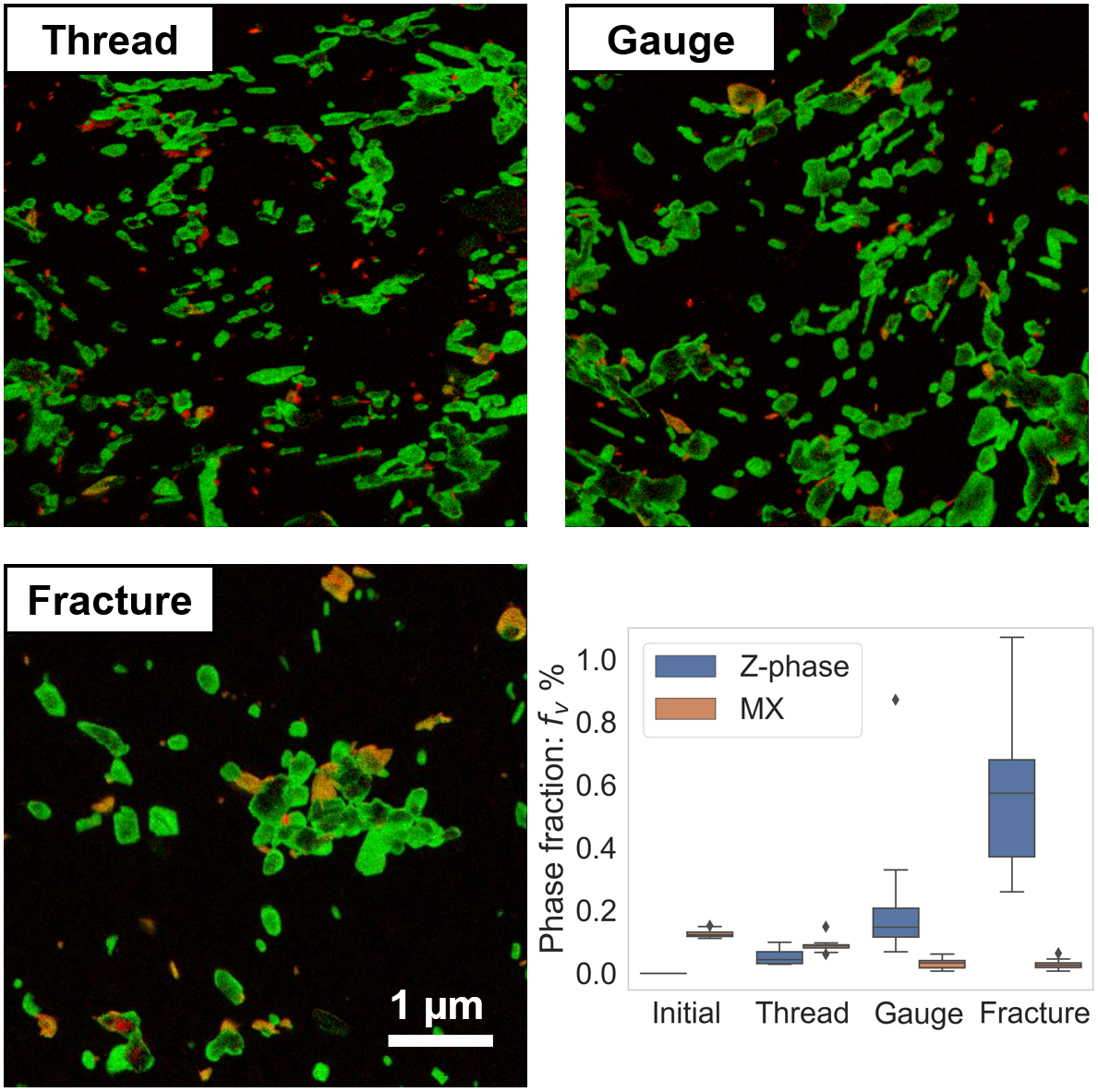}
  \end{center}
  \caption{EFTEM elemental maps shown as a colour overlay (Cr:green; V:red) for the different regions of the ruptured sample, consisting of the \emph{Thread}, \emph{Gauge} and \emph{Fracture} area, where the localised creep-strain resulted in rupture of the creep specimen.}
  \label{fig:EFTEM_RGB_Replica_Fracture}
  \end{figure}

In order to investigate the effects of creep-strain, applied stress and ageing temperature on the formation of modified \emph{Z}-phase and the associated dissolution of the MX precipitates, the different sections (\emph{Fracture}; \emph{Gauge}; \emph{Thread}) of the ruptured creep-tested sample (139 kh) were compared to the initial state. Figure \ref{fig:EFTEM_RGB_Replica_Fracture} shows the RGB EFTEM maps taken from extraction replicas prepared from the different locations of the ruptured sample.  At the location of fracture the test specimen underwent a significant reduction in cross-section area (approximately 50\% for a diameter decrease from 6.0 mm to 4.2 mm) and most of the creep-strain (11.9\%) would have been localised in this region. FIB-lamellae (5x5 $\mu m^2$) were prepared from ten random sites very close to the fracture surface and an extraction replica sample was also prepared from this region. Quantitative analysis of the modified \emph{Z}-phase and MX precipitates was performed on the extraction replicas as previously described. The phase fractions of MX and modified \emph{Z}-phase are shown as an insert to Figure\ref{fig:EFTEM_RGB_Replica_Fracture}.

The thread region still had numerous MX precipitates compared to the low number of MX carbonitride particles observed in the (uniformly-elongated) gauge region and the fracture region. Additionally, the projected area number density of precipitates is much lower for the extraction replica prepared from the fracture region. This is probably due to the significant microstructural coarsening that occurred during the final stages of tertiary creep, prior to fracture. Consequently, the calculated sampled depth for the extraction replica prepared from the fracture region was only $0.5 \mu m$ compared to $1 \mu m$ for the extraction replicas prepared from the other material states. The phase fraction of modified \emph{Z}-phase in the fracture region ($f_v$ = 0.40 $\pm$ 0.20 \%) is significantly higher than in the uniformly-elongated gauge section. This implies that the localised creep-strain is correlated with modified \emph{Z}-phase formation. 

\subsection{Preferential nucleation sites for modified \emph{Z}-phase formation}

\begin{figure}
  \begin{center}
    \includegraphics[width=0.45\textwidth]{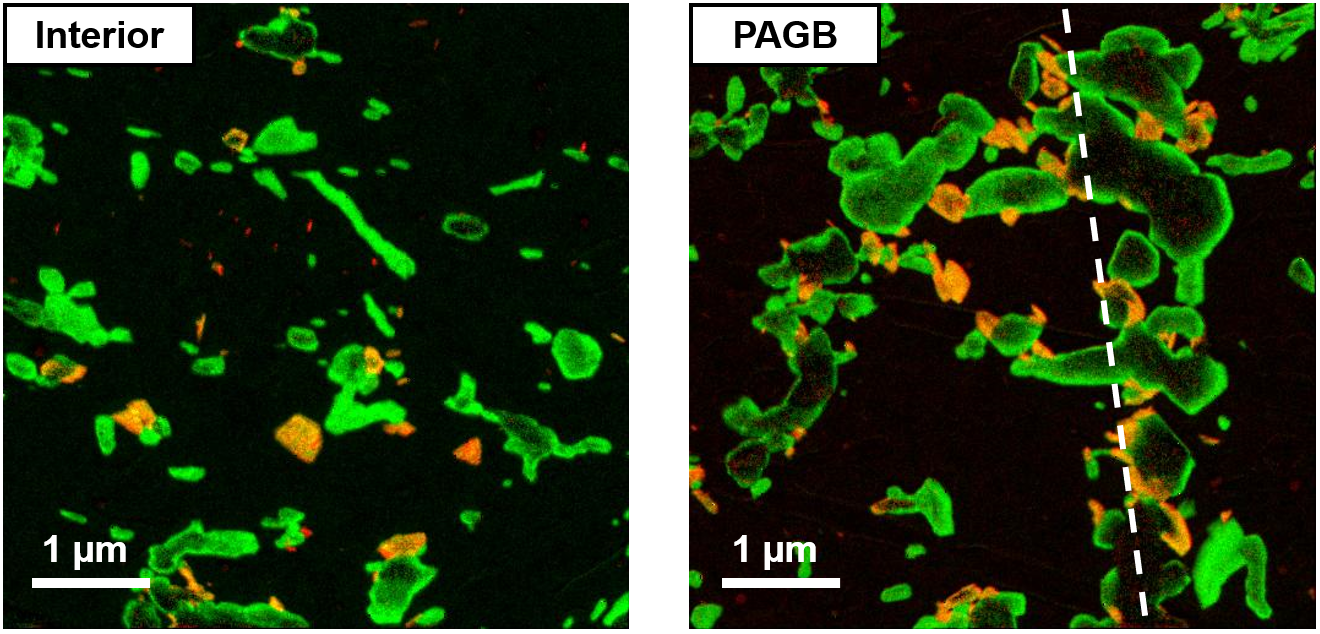}
  \end{center}
  \caption{Typical EFTEM elemental maps shown as a colour overlay (Cr:green; V:red) for  the extraction replica prepared from the gauge region of the ruptured sample, showing the PAGB interior (left) and the PAGB (indicated by a white broken line on the right).}
  \label{fig:EFTEM_RGB_ZPhase_Location}
  \end{figure}

The analysis was then performed at \emph{prior austenite grain interiors} and at \emph{prior-austenite boundaries} on the extraction replica prepared from the gauge region of the ruptured sample. Figure \ref{fig:EFTEM_RGB_ZPhase_Location} shows two representative EFTEM elemental maps for the two different locations of a prior-austenite grain. When viewed in projection, it appears that the PAGB locations contain more modified \emph{Z}-phase particles, however, the PAGBs had a significantly larger number density of M\textsubscript{23}C\textsubscript{6} precipitates in a given field of view, which resulted in a larger extraction depth estimation. The values of the modified \emph{Z}-phase volumetric phase fraction ($f_v$) are not significantly different for the two locations when the correction is made for the extraction depth. This remains a limitation for quantitative analysis performed on 5x5 $\mu m^2$ areas of an extraction replica and it could not be confirmed whether or not the prior-austenite grain boundaries promote the formation of the modified \emph{Z}-phase. 

\begin{figure}
  \begin{center}
    \includegraphics[width=0.45\textwidth]{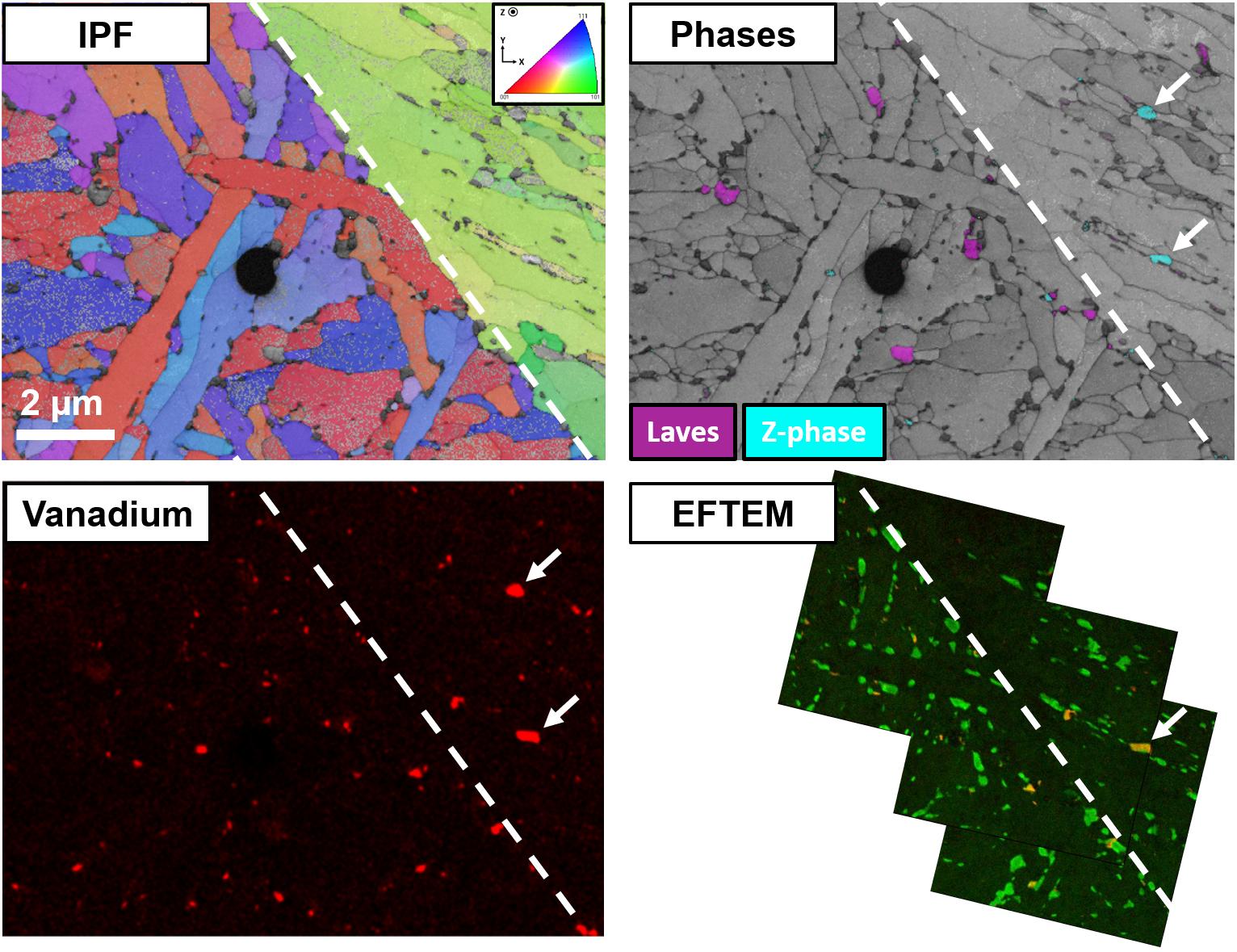}
  \end{center}
  \caption{Combined TKD-EDS maps of over a PAGB in a thin-foil sample prepared from the gauge area of the ruptured sample. The lower right image is an EFTEM composite elemental map taken from the exact same area.}
  \label{fig:TKD_Foil}
  \end{figure}

  TKD-EDS analyses were performed on the thin-foil sample prepared from the uniformly-elongated gauge region of the sample tested to rupture. The thin-foil was scanned with a step size of 20 $nm$ across two prior-austenite grains, which was confirmed by employing a reconstruction algorithm (results not shown). Figure \ref{fig:TKD_Foil} shows the results of the scan. Several vanadium-rich precipitates can be seen in the EDS map. Two of the larger precipitate phases (white arrows) indexed successfully as the modified \emph{Z}-phase precipitates with a tetragonal crystal structure. However, they were not located exactly on the PAGB or a packet boundary. The distribution of vanadium-rich precipitates, which did not index successfully, were also not preferentially aligned along the PAGB. EFTEM elemental mapping of this area is shown for comparison. 

\section{Discussion}

This study investigated the formation of modified \emph{Z}-phase and the associated dissolution of creep-strengthening MX precipitates in a 12\%Cr (German-grade) X20 TMF steel using quantitative electron microscopy. The previous investigations by Aghajani \emph{et. al} \citep{aghajani2009b} used HAADF-STEM imaging and EDS mapping on thin-foils to identify the MX precipitates, and found the phase fraction of MX precipitates to be constant until rupture and they did not find the modified \emph{Z}-phase. The current study used EDS detectors with an improved collection efficiency, employed more sensitive EFTEM elemental analysis and performed measurements on extraction replicas for a more representative sampling volume. Clear evidence (crystallographic and chemical) of modified \emph{Z}-phase was observed after a creep-test duration of 51 kh in the uniformly-elongated gauge section. The phase fraction ($f_v$) and size ($d_m$) of the modified \emph{Z}-phase increased with increasing testing time, with an associated decrease in the phase fraction ($f_v$) and number density ($N_v$) of the MX particles for the samples in the uniformly-elongated gauge sections. This resulted in an increase in the inter-particle distance for the MX carbonitrides ($\lambda_{3D}$) as compared to the initial state material.

\subsection{Quantitative characterisation methods}
Several quantitative microscopy techniques were used in this study. Extraction replication is a relatively simple sample preparation technique, which allows for investigations on large areas without the influence of the magnetic iron matrix. However, the extraction volume for this preparation method is unknown. We provided a way to estimate this volume and used it quantify the volumetric phase fractions of MX particles based on EFTEM analysis. Previous investigations \citep{danielsen2006a, sawada2006, cipolla2010, sawada2014} used projected area number density ($N_A$) as a quantitative measure. However, variations in sample preparation procedures and the microstructural state can strongly influence the sampling depth from which the precipitates are extracted. The method proposed in this study attempts to correct for this variation in the extraction depth for the replica samples. The measured values of the volumetric phase fraction for MX carbonitrides are systematically lower for the extraction replica samples. This could be due to lower extraction efficiency of smaller particles or systematic over-estimation of the extraction volume, due to approximations in the method that was employed to determine the sampled volume from the extraction replica samples. 

Thin-foils preserve the precipitate locations and the investigated volume can be accurately determined by measuring the thickness of the foil with EELS. However, the precipitate feature measurements have to be stereologically corrected to calculate the volumetric measurements, due to the fact that the precipitates are sectioned and then viewed as a projected area. The sampled volume for thin-foils is approximately 10\% of that of the extraction replica method for a given site of interest. It is important to perform measurements from several sites of interest per material state and then to calculate the mean value for each area. The standard deviation in the mean values will be an indication of the representative nature of the measurements for a particular microstructural feature. Combined TKD-EDS SEM analyses performed on the thin-foils were able to map out the precipitate species with a spatial resolution in the order of 20 nm to obtain crystallographic phase, orientation and compositional information.  

The modified \emph{Z}-phase particles were quantified from RGB EFTEM elemental maps. Image segmentation was performed to identify the modified \emph{Z}-phase based on the intensity of the chromium and vanadium signals. It is very difficult to distinguish between a modified \emph{Z}-phase particle and an MX/M\textsubscript{23}C\textsubscript{6} overlap for the early stages (12 kh). Segmentation errors are an important consideration that could influence the quantitative results in this study and it must be performed with care. Modified \emph{Z}-phase particles identified in the uniformly-elongated gauge section of the sample creep-tested to 12 kh should be interpreted with care, since these particles almost exclusively occurred adjacent to M\textsubscript{23}C\textsubscript{6} precipitates and could be due to MX/M\textsubscript{23}C\textsubscript{6} overlapping Cr/V signals. These phases could not be identified as the modified \emph{Z}-phase using electron diffraction due to their small size and close proximity to the M\textsubscript{23}C\textsubscript{6} particles.

\subsection{Effects of creep-strain on modified Z-phase formation} Characterisation of extraction replicas by Sawada \emph{et. al} \citep{sawada2006, sawada2014} found that the gauge sections contained 2-4 times the phase fraction of modified \emph{Z}-phases as compared to the thread regions of the \emph{creep-ruptured} samples. The phase fraction of modified \emph{Z}-phase of the deformed fracture region, in this study, was significantly higher ($f_v$: 0.40 $\pm$ 0.02 \%) compared to uniformly-elongated zone ($f_v$: 0.23 $\pm$ 0.20 \%) of the sample tested to rupture. This suggests that localised creep-strain and increased stress due to necking promoted the formation of modified \emph{Z}-phase. Recent thermo-kinetical modeling by Svoboda \citep{svoboda2019} suggests that deformation processes may relax the volumetric back-pressure created during the formation of modified \emph{Z}-phase, but this is difficult to model accurately. If localised creep-strain enhances the formation of the modified \emph{Z}-phase, then it could possibly explain the low number density of modified \emph{Z}-phase particles observed by Danielsen \citep{danielsen2006} in the service-exposed X20 sample that was probably not subjected to significant levels of localised creep-strain.  

\subsection{Influence of modified Z-phase on creep-strength}
It is quite challenging to quantify the contribution of the MX precipitate dissolution towards the progressive loss of creep-resistance, since it is also influenced by a combination of several microstructural mechanisms as discussed by Aghajani \emph{et. al} \citep{aghajani2009b}. Recent developments in microstructurally based semi-physical models \citep{riedlsperger2020a} use quantitative microstructural data directly to calculate the time-to-rupture diagrams and can estimate the effects of individual microstructural features on the creep-resistance. The effects of MX dissolution on the creep-resistance using this approach will be the subject of a follow-up study, since it can't be quantitatively described without taking the evolution of all the microstructural features (including sub-grains, M\textsubscript{23}C\textsubscript{6}, Laves, and dislocations) during creep-testing into account \citep{holzer2010}.

\subsection{Nucleation sites for modified Z-phase}
Previous investigations found that the modified \emph{Z}-phase formed preferentially on prior-austenite and packet boundaries \citep{sawada2006, sawada2014, danielsen2006a}. Extraction replicas were used and it was assumed that the lines of enlarged M\textsubscript{23}C\textsubscript{6} precipitates delineate the prior-austenite grain boundaries, since the iron grain information is lost during the extraction process. The projected area number density ($N_A$) of the precipitates from the extraction replica sample was presented as a quantitative measure in those investigations, but the extraction depth is very dependent on the sample preparation and the underlying microstructural grain-size as previously discussed. 

The quantitative investigations performed in this study to determine the preferential nucleation sites for the modified \emph{Z}-phase, could not conclusively show whether modified \emph{Z}-phase forms preferentially on the PAGB in this X20 steel grade, since several particles were distributed throughout the grain structure away from the PAGB. Qualitative evaluation found that most of the modified \emph{Z}-phase particles were located adjacent to the M\textsubscript{23}C\textsubscript{6} precipitates and the modified \emph{Z}-phase formed through the transformation of MX particles by chromium diffusion. The modified \emph{Z}-phase number density remains constant ($N_v$ = 3 $\pm$ 1 $\mu m^{-3}$) once formed and only increases in size by further dissolution of MX particles through the matrix diffusion of vanadium and chromium. A detailed study of the preferential nucleation sites of the modified \emph{Z}-phase will be the subject of a follow-up study using TKD-EDS, HR(S)TEM-EELS, and atom-probe tomography.   


\section{Conclusions}
1. Modified \emph{Z}-phase was observed in German-grade (X20) material after creep-testing for 51 kh (120 MPa at \SI{550}{\celsius}) in the uniformly-elongated gauge portion of the sample. This phase increased in size with an accompanied dissolution of MX particles until rupture after 139 kh.

2. The effects of stress and localised creep-strain significantly increased the formation of the modified \emph{Z}-phase for the sample tested to failure. 

3. The modified \emph{Z}-phase (CrVN) formed preferentially near M\textsubscript{23}C\textsubscript{6} precipitates and no conclusive evidence for preferential formation at PAGBs could be found based on investigations performed on the uniformly-elongated gauge section of the ruptured X20 grade sample.

4. The modified \emph{Z}-phase formation and dissolution of MX is considered to be an important microstructural degradation mechanism that results in a non-linear decrease in creep-strength for 11-12\% Cr TMF steels. The microstructural characterisation techniques demonstrated in this study can be applied to quantitatively characterise the finest microstructural features (MX) which are considered critical to the creep-strength of 9-12\% TMF steel grades. 

\section*{Data availability}
The raw data required to reproduce these finding are available to download after a request to the corresponding author. 

\section*{Declaration of Competing Interest}
The authors declare that they have no known competing financial interests or personal relationships that could have appeared to influence the work reported in this paper. 

\section*{\emph{Acknowledgement}}
JEW and WEG gratefully acknowledge financial support of the National Research Foundation of South Africa (Grant number 70724) and the Materials and Mechanics Specialisation Group at the University of Cape Town through the Eskom Power Plant Engineering Institute (EPPEI). 
HW, AK and GE acknowledge support from the Centre for Interface Dominated High Performance Materials (German abbreviation: ZGH)

\bibliography{GOO220_A1}

\end{document}